\documentclass[aps,prl,superscriptaddress,twoside,twocolumn,nofootinbib,preprintnumbers]{revtex4}

\usepackage{amsmath}
\usepackage{graphicx}

\begin{document}

\title{Conformal Barrier for New Vector Bosons Decay to the Higgs}  

\author{Hidenori S. Fukano}\thanks{{\tt fukano@kmi.nagoya-u.ac.jp}}
      \affiliation{ Kobayashi-Maskawa Institute for the Origin of Particles and 
the Universe (KMI) \\ 
 Nagoya University, Nagoya 464-8602, Japan.}
\author{Shinya Matsuzaki}\thanks{{\tt synya@hken.phys.nagoya-u.ac.jp}}
      \affiliation{ Institute for Advanced Research, Nagoya University, Nagoya 464-8602, Japan.}
      \affiliation{ Department of Physics, Nagoya University, Nagoya 464-8602, Japan.}    
\author{Koichi Yamawaki} \thanks{{\tt yamawaki@kmi.nagoya-u.ac.jp}}
      \affiliation{ Kobayashi-Maskawa Institute for the Origin of Particles and the Universe (KMI) \\ 
 Nagoya University, Nagoya 464-8602, Japan.}

\date{\today}

\begin{abstract} 
The ATLAS collaboration has recently reported excesses about 2.5 sigma 
at mass around 2 TeV in the diboson channels, which can be identified with new vector bosons 
as a hint for the new physics. 
It is shown 
that spontaneously broken conformal/scale symmetry prohibits new vector bosons
decay to the Higgs, 
which is contrasted  to the popular 
``equivalence theorem" valid only in a special limit not necessarily relevant to the 2 TeV mass.  
If the decay $V\to WH/ZH$ is not observed in
the ongoing Run II of the LHC, then the 125 GeV Higgs can be a dilaton.
\end{abstract}
\maketitle


After the Higgs boson was discovered at LHC, there
have been detailed LHC analyses of the Higgs, which is
consistent with the standard model so far~\cite{Aad:2012tfa,Chatrchyan:2012ufa}, 
without serious hints for the new physics beyond the standard
model. However, the origin of the mass of the Higgs itself is still a 
biggest mystery of modern particle physics, 
which would imply new physics beyond the standard model.

Very recently, the ATLAS collaboration~\cite{Aad:2015owa} has reported
excesses about 2.5 sigma (at global significance)
with narrow width less than 100 GeV at mass around 2
TeV in the diboson channels~\footnote{
Small excesses about $\sim 2$ sigma in the same mass region have been seen 
also in the CMS diboson analysis~\cite{Khachatryan:2014hpa}. }. 
If it is confirmed in the LHC Run II, it will certainly be  
an outstanding signature of new physics. 
It should be deeply connected with the
long-standing mystery, such as the naturalness, of the dynamical
origin of the Higgs itself. Hence the events not
only are exciting in their own right but also would be important
to giving important clues to understand the nature of the Higgs.

With such excitements, the diboson events have already
attracted a lot of attention proposing possible candidates
for the origin of the excess, 
such as a new vector boson ($V$) like technirhos~\cite{Fukano:2015hga,Franzosi:2015zra}, 
$W'/Z'$~\cite{Hisano:2015gna,Cheung:2015nha,Dobrescu:2015qna,Aguilar-Saavedra:2015rna,Alves:2015mua,Gao:2015irw,Thamm:2015csa,Brehmer:2015cia,Cao:2015lia,Cacciapaglia:2015eea,Abe:2015jra,Abe:2015uaa,Carmona:2015xaa}, or others~\cite{Chiang:2015lqa}.     
Most of such 2 TeV vector resonance models involves the vector boson decays to 
weak boson pairs ($WW, WZ$), 
as well as the decays along with the 125 GeV Higgs ($WH, ZH$). 
The ratio of the two decay rates 
is almost one, $\Gamma(V \to WW/WZ)/\Gamma(V \to WH/ZH) \simeq 1$, 
according to the 
popular ``equivalence theorem'', see e.g., \cite{Hisano:2015gna}. 
Hence one naively expects to discover the $V$ not only in the $WW/WZ$ channels, but also 
in the $WH/ZH$ channels.  
Therefore the
present CMS experimental bounds~\cite{Aad:2015yza,Khachatryan:2015bma} on the latter processes
have already given stringent constraints on the 
generic vector models.

In this Letter we propose a novel way to identify the dynamical 
origin of the 125 GeV Higgs through checking the possible
decays of the 2 TeV new bosons.   
If the 2 TeV new
bosons have no decays to the SM gauge bosons plus the 125
GeV Higgs 
then we show that the 125 GeV Higgs can be a dilaton, 
pseudo Nambu-Goldstone boson of the 
spontaneously broken conformality/scale symmetry of some
underlying new physics, with the scale symmetry broken
also explicitly 
only by the Higgs mass in the effective theory.
One such an explicit example of the underlying theory
is the walking technicolor~\cite{Yamawaki:1985zg}
where the 125 GeV Higgs 
and the new bosons have 
been successfully identified with the technidilaton~\cite{Matsuzaki:2012gd,Matsuzaki:2012vc,Matsuzaki:2012mk,Matsuzaki:2013fqa,Matsuzaki:2012xx,Shinya:scgt15} 
and the walking technirho~\cite{Fukano:2015hga}, respectively.

We begin with a generic model, called heavy-vector triplet (HVT) model~\cite{Pappadopulo:2014qza},     
which is quoted by the ATLAS and CMS groups for new vector boson searches as a benchmark.  
The model Lagrangian reads~\cite{Pappadopulo:2014qza} 
\begin{eqnarray} 
 {\cal L}_V 
&=&  
- \frac{1}{2} {\rm tr}[ V_{\mu\nu}^2] + m_V^2 {\rm tr}[V_\mu^2]  
\nonumber \\ 
&& 
+ g_V c_{_{H}} \, \left( i H^\dag V^\mu D_\mu H + {\rm h.c.} \right)
\nonumber \\ 
&& 
+ 2 g_V^2 c_{_{VVHH}} {\rm tr}[ V_\mu^2 ] H^\dag H 
\nonumber \\ 
&& 
+
{\cal L}_{\rm Higgs} 
+  \cdots 
\,, \label{LV} 
\end{eqnarray}
where we have put the standard-model Higgs terms ${\cal L}_{\rm Higgs}$ including 
the kinetic term $|D_\mu H|^2$ and the usual Higgs potential ($V_{\rm Higgs}$). 
In Eq.(\ref{LV}) we have defined 
$V_{\mu\nu} = D_\mu V_\nu - D_\nu V_\mu$    
with $D_\mu V_\nu = \partial_\mu V_\nu - i g_W [W_\mu, V_\nu]$ with the $g_W$ being the weak gauge couping.    
and have not displayed terms which do not include the Higgs $H$ along with the new vector boson field $V$.

  When the Higgs field $H$ gets the vacuum expectation value $v$ $(\simeq$ 246 GeV), 
the new vector boson $V$ starts to mix with the weak boson $W$ through the $c_{_{H}}$ term in Eq.(\ref{LV}).  
Parameterizing the $H$ as $H= v/\sqrt{2} (1 + \phi/v) (0,1)^T$ plus the eaten Nambu-Goldstone boson terms 
and ignoring the hypercharge gauge for simplicity,   
one finds the mass matrix for ${\bf V}_\mu$ = ($V_\mu$, $W_\mu$)$^T$, 
\begin{equation}  
{\cal M}^2 
= 
 \left( 
\begin{array}{cc} 
m_V^2 + g_V^2 c_{_{VVHH}} v^2 &  \frac{1}{4} g_W g_V c_{_{H}} v^2 \\ 
  \frac{1}{4} g_W g_V c_{_{H}} v^2 & \frac{1}{4} g_W^2 v^2 
\end{array} 
\right) 
\,. \label{mass-matrix}
\end{equation}  
In addition, one has the Higgs ($\phi$) couplings to $V$ and $W$, 
\begin{equation} 
{\cal G}_{VW\phi} 
= 
 \left( 
\begin{array}{cc} 
g_V^2 c_{_{VVHH}} v^2 &  \frac{1}{4} g_W g_V c_{_{H}} v^2 \\ 
  \frac{1}{4} g_W g_V c_{_{H}} v^2 & \frac{1}{4} g_W^2 v^2 
\end{array} 
\right) 
\,. \label{VHW}
\end{equation} 
In the Lagrangian the ${\cal M}^2 $ and ${\cal G}_{VW\phi} $ terms look like 
\begin{equation} 
 {\cal L}_V = \frac{1}{2} 
{\bf V}^T_\mu \cdot {\cal M}^2 \cdot {\bf V}^\mu 
+ 
\frac{\phi}{v} \cdot {\bf V}_\mu^T \cdot {\cal G}_{VW\phi} \cdot {\bf V}^\mu 
- V_{\rm Higgs} + \cdots 
\,.   
\end{equation} 
Note that the mass matrix ${\cal M}^2$ and the couplings to the Higgs $\phi$ differ 
only by the $m_V^2$ term.  
After diagonalizing the mass matrix Eq.(\ref{mass-matrix}), one gets the 
mass eigenstates $\tilde{\bf V} = (\tilde{V}, \tilde{W})$ and finds the couplings such as 
$\tilde{V}$-$\tilde{V}$-$\phi$, $\tilde{W}$-$\tilde{W}$-$\phi$, as well as the off diagonal 
coupling $\tilde{V}$-$\tilde{W}$-$\phi$. 
The presence of the nonzero off-diagonal coupling $\tilde{V}$-$\tilde{W}$-$\phi$ 
is essentially due to the $m_V^2$ term in Eq.(\ref{LV}): without the $m_V^2$ term 
two mixing matrices ${\cal M}^2$ and ${\cal G}_{VW \phi}$ would become identical to be diagonalized simultaneously, 
so the $\tilde{V}$-$\tilde{W}$-$\phi$ coupling would completely be rotated away.

Now, we shall introduce the conformal/scale invariance into the HVT model in Eq.(\ref{LV}).  
Examining terms in Eq.(\ref{LV}) in quadratic order of the vector fields with the scale dimensions taken into account,  
one readily realizes that only the $m_V^2$ term violates the scale invariance 
for the action corresponding to the model Lagrangian Eq.(\ref{LV})~\footnote{
Of course, the scale invariance would be broken at the loop level, 
as will be addressed below. }. 
Absence of this term does not affect 2 TeV mass of the new boson. 
Eliminating the $m_V^2$ term, the conformal/scale invariance thus leads to 
the mass matrix 
\begin{equation} 
{\cal M}^2_{m_V=0}  
= 
 \left( 
\begin{array}{cc} 
g_V^2 c_{_{VVHH}} v^2 &  \frac{1}{4} g_W g_V c_{_{H}} v^2 \\ 
  \frac{1}{4} g_W g_V c_{_{H}} v^2 & \frac{1}{4} g_W^2 v^2 
\end{array} 
\right) 
\,. \label{mass-matrix:C}
\end{equation}
This is the same matrix as the ${\cal G}_{VW\phi}$ in Eq.(\ref{VHW}), 
hence the off-diagonal $\tilde{V}$-$\tilde{W}$-$\phi$ coupling goes away after the diagonalization of the vector boson 
sector: 
\begin{eqnarray} 
{\cal L}_V \Bigg|_{m_V=0} 
&=& 
\frac{1}{2} 
{\bf V}^T_\mu \cdot {\cal M}^2_{m_V=0} \cdot {\bf V}^\mu 
+ 
\frac{\phi}{v} \cdot {\bf V}_\mu^T \cdot {\cal G}_{VW\phi} \cdot {\bf V}^\mu 
\nonumber \\ 
&& 
+ 
\cdots 
\nonumber \\ 
&=& 
\frac{1}{2} 
\left(1 + \frac{2 \phi}{v} \right) 
{\bf V}^T_\mu \cdot 
{\cal M}^2_{m_V=0} 
\cdot 
{\bf V}^\mu 
\nonumber \\ 
&& 
+ \cdots 
\,. \label{LV:C}
\end{eqnarray} 
In terms of the mass eigenstate fields $\tilde{\bf V}_\mu = (\tilde{V}_\mu, \tilde{W}_\mu)^T$, 
the Lagrangian Eq.(\ref{LV:C}) goes like 
\begin{eqnarray} 
 {\cal L}_V \Bigg|_{m_V=0} 
& =& 
 \frac{1}{2} 
\left(1 + \frac{2 \phi}{v} \right) 
\tilde{\bf V}^T_\mu \cdot 
 \left( 
\begin{array}{cc} 
m_{\tilde V}^2 &  0 \\ 
  0 & m_{\tilde{W}}^2 
\end{array} 
\right)   
\cdot 
\tilde{\bf V}^\mu 
\nonumber \\ 
&& 
+ \cdots 
\,, 
\end{eqnarray}
with the masses of the mass eigenstate vectors $(m_{\tilde{V}}, m_{\tilde W})$.


The new vector boson $V$ thus does not decay to the weak bosons 
in association with the Higgs 
in the presence of the scale/conformal symmetry ({\it Conformal Barrier}), i.e., 
\begin{equation} 
  V - W/Z - H \, {\rm coupling} \, = \, 0 
  \,, 
\end{equation} 
consequently the $V$ predominantly decays to the weak boson pairs $WW/WZ$. 
The absence of $V\to WH/ZH$ signatures at the LHC Run-II could indirectly probe the existence of the (approximate) scale/conformal invariance.

The conformal/scale-invariant limit ($m_V^2 \to 0$ in Eq.(\ref{mass-matrix})) 
with the strong coupling ($g_V \gg 1$) is perfectly consistent with the mass $m_{\tilde V} \simeq g_V v \simeq 2$ TeV~\footnote{
It implies a large coupling $g_V \sim 10$, which would lead to sizable corrections 
through the new vector boson loops to other couplings, say, 
higgs self-couplings. The size of such loop corrections would be quite   
large (${\cal O}(g_V^4/(4\pi)^2)={\cal O}(10^2)$), implying 
that the naive perturbation in $g_V$ would break down, which needs 
some ultraviolet completion like walking technicolor. 
\label{UV-comp}}, in a way 
incompatible with the so-called ``equivalence theorem" for the $V \to WW/WZ$ and $V \to WH/ZH$ 
decays, i.e., $\Gamma(V \to WW/WZ) \simeq \Gamma(V \to WH/ZH)$, which actually can only be achieved  by 
taking a special limit $m_V \gg g_V v \,\,(\gg g_W v)$.

The conformal/scale invariance should be approximate, hence the conformal barrier will 
be broken at higher order level of the perturbation theory. 
If the symmetry is explicitly broken only by the Higgs mass term 
$\frac{1}{2} m_\phi^2 \phi^2$ (soft-breaking)~\footnote{
Anther explicit breaking for the scale symmetry would arise 
as the usual trace anomaly term like the Higgs-diphoton coupling of 
$\phi F_{\mu\nu}^2$ form.    
Since the new vector boson mass arises only 
from the electroweak scale $v$ in the presence of the conformal barrier ($m_V=0$), 
the charged new vector boson would then contribute to $\phi F_{\mu\nu}^2$ 
as a nondecoupling effect, to be strongly constrained by current 
precise measurements of the Higgs-diphoton coupling at LHC.  
However, 
the vector boson loop corrections would be nonperturbative 
because of the large coupling $g_V \sim 10$ (See also footnote~\ref{UV-comp}), 
so that some ultraviolet completion is needed to properly estimate 
the size of the corrections, as done in the scenario of the walking technicolor.  
},  
then the trilinear Higgs coupling proportional to the Higgs mass would give rise to 
the $\tilde{V}$-$\tilde{W}$-$\phi$ at the two loop level, which is, however, 
too tiny to be detected at the LHC experiments.

To see the conformal/scale invariance  more manifestly, 
we may rewrite the Lagrangian ${\cal L}_V|_{m_V=0}$ in Eq.(\ref{LV:C}) 
into the nonlinear realization for the conformal/scale symmetry by 
introducing the nonlinear base $\chi = e^{\phi/v} = 1 + \phi/v + \cdots$ as   
\begin{eqnarray} 
{\cal L}_V \Bigg|_{m_V=0} 
& = & 
\frac{1}{2} 
\chi^2  
{\bf V}^T_\mu 
 \left( 
\begin{array}{cc} 
g_V^2 c_{_{VVHH}} v^2 &  \frac{1}{4} g_W g_V c_{_{H}} v^2 \\ 
  \frac{1}{4} g_W g_V c_{_{H}} v^2 & \frac{1}{4} g_W^2 v^2 
\end{array} 
\right)   
{\bf V}^\mu 
\nonumber \\ 
&& 
+ \cdots 
\,. 
\label{L:sHVT}
\end{eqnarray}  
The form of this Lagrangian implies that the Higgs $\phi$ 
is nothing but a dilaton,  
transforming as $\delta \phi(x) = (v + x^\nu \partial_\nu \phi(x))$ and 
$\delta \chi(x) = (1 + x_\nu \partial^\nu) \chi(x)$, where 
the $v$ is identified with the dilaton decay constant $F_\phi = v$.

 Actually, the decay constant of the dilaton is not necessarily 
 equal to the $v$ ($\chi = e^{\phi/F_\phi}$ with $F_\phi \neq v$): 
the most general vector boson action invariant under the conformal/scale invariance 
is given by the scale-invariant version of the hidden local 
symmetry (sHLS) Lagrangian~\cite{Bando:1984ej,Bando:1985rf,Bando:1988ym,Bando:1988br}, 
\begin{equation} 
 {\cal L}_{\rm sHLS} 
 = \chi^2 F_\pi^2 \left( 
{\rm tr}[ \hat{\alpha}_{\mu \perp}^2 ] 
+ 
a \, {\rm tr}[ \hat{\alpha}_{\mu ||}^2 ] 
\right) 
+ \cdots  
\,, \label{sHLS}
\end{equation}     
where $\hat{\alpha}_{\perp, ||} = (D_\mu \xi_R \xi_R^\dag \mp D_\mu \xi_L \xi_L^\dag)/(2i)$ 
with $D_\mu \xi_{R,L} = \partial_\mu \xi_{R,L} - i V_\mu \xi_{R,L} 
+ i  \xi_{R,L} {\cal R}_\mu({\cal L}_\mu)$. 
The nonlinear bases $\xi_R$ and $\xi_L$ for the chiral $SU(N_F)_L \times SU(N)_R$ 
symmetry form the chiral field $U$ as $U=\xi_L^\dag \xi_R$, which transforms 
as $U \to g_L \cdot U \cdot g_R^\dag$ with the electroweak gauges partially 
embedded in 
$g_L$ and $g_R$ as well as the standard-model gauge bosons in the gauge fields 
${\cal L}_\mu$ and ${\cal R}_\mu$. 
The new vector bosons ($V_\mu$) have been introduced as gauge bosons of the HLS.       
The decay constant $F_\pi$ is related to 
the electroweak scale $v$ as $F_\pi^2 = v^2/(N_F/2)$ and the arbitrary parameter $a$ 
can be phenomenologically fixed.

 To make a direct comparison with the scale-invariant HVT model in Eq.(\ref{L:sHVT}), 
we shall take $N_F=2$ and expand the sHLS Lagrangian to get 
the mass matrix for the electroweak bosons $W_\mu$ 
and the new vector bosons $V_\mu$ (${\bf V}_\mu = (V_\mu, W_\mu)^T)$: 
\begin{eqnarray} 
{\cal L}_{\rm sHLS} 
&=& 
\frac{1}{2} 
\chi^2  
{\bf V}^T_\mu 
 \left( 
\begin{array}{cc} 
a \,  g^2 \, v^2 
 & - \frac{a}{2} g\,  g_W \, v^2 \\ 
  - \frac{a}{2} g\,  g_W \, v^2 &  \frac{(1+ a)}{4}  g_W^2 v^2  
\end{array} 
\right)   
{\bf V}^\mu 
\nonumber \\ 
&& 
+ 
\cdots 
\,, \label{L:sHLS}
\end{eqnarray}
where we have introduced the new-vector boson 
kinetic term $( - \frac{1}{2g^2} {\rm tr}[V_{\mu\nu}^2])$ 
with the gauge coupling $g$ and rescaled the vector fields canonically.  
It is obvious that the mass matrix and the vertices involving the Higgs = dilaton $\phi$ 
are simultaneously diagonalized away in the same way as in 
Eq.(\ref{L:sHLS}): the conformal/scale symmetry prohibits the new vector boson $V$ from 
decaying to the Higgs.

Conversely, if the decay of new vector bosons into the Higgs 
is not observed in the ongoing Run II of the LHC, then 
it is suggested that the Higgs is a dilaton.

The sHLS Lagrangian in Eq.(\ref{sHLS}) is the effective theory  
realizing the (approximate) scale/conformal invariance and chiral symmetry of 
the underlying theory, the walking technicolor~\cite{Yamawaki:1985zg}. 
In the walking technicolor, the Higgs is nothing but the technidilaton ($\phi$), 
a composite pseudo Nambu-Goldstone boson for the (approximate) conformal/scale symmetry,  
and the new vector bosons are the technirhos ($V$). 
The conformal/scale symmetry of the sHLS is explicitly broken by 
the dilaton mass in the potential of the 
form $\sim$ $F_\phi^2 m_\phi^2 \chi^4 (\log \chi -1/4 )$, which corresponds to 
the trace anomaly of the underlying walking technicolor. 
The effect of the symmetry breaking arises only at ${\cal O}(p^6)$ or higher orders 
of the derivative expansion, since the ${\cal O}(p^4)$ terms are 
already scale-invariant without involving 
the technidilaton field $\chi = e^{\phi/F_\phi}$. 
Thus, additional Higgs (= $\phi$) potential terms are not generated at the ${\cal O}(p^4)$.      
Though the dilaton decay constant $F_\phi$ is in principle determined from 
the walking dynamics itself, the value of $F_\phi$ can be fitted 
to the LHC Higgs coupling data~\cite{Matsuzaki:2012gd,Matsuzaki:2012vc,Matsuzaki:2012mk,Matsuzaki:2013fqa,Matsuzaki:2012xx,Shinya:scgt15}, while 
the diboson channels are totally blind against the $F_\phi$ because of 
the absence of the $V \to W\phi/Z\phi$ modes~\footnote{ 
In this respect, 
the analysis in Ref.~\cite{Kurachi:2014qma} is subject to modifications,   
which will be given in another communication. 
Especially, there are no couplings between $\phi$, gluon $g$ and color-octet technirhos. }.

 In Ref.~\cite{Fukano:2015hga} it is shown that 
 the 2 TeV technirho can explain the ATLAS diboson excesses. 
Although in Ref.~\cite{Fukano:2015hga}  
the one-family model is taken as a realistic walking technicolor, 
the actual analysis of diboson events is free from 
the model-dependent parameters $a$ and $F_\phi$. 
The absence of the $W\phi$ and $Z\phi$ channels thus leads to the significantly large 
$V \to WW/WZ$ cross sections, compared to other types of vector bosons (e.g., $W'/Z'$) 
without the scale invariance~\cite{Franzosi:2015zra,Hisano:2015gna,Cheung:2015nha,Dobrescu:2015qna,Aguilar-Saavedra:2015rna,Alves:2015mua,Gao:2015irw,Thamm:2015csa,Brehmer:2015cia,Cao:2015lia,Cacciapaglia:2015eea,Abe:2015jra,Abe:2015uaa,Carmona:2015xaa}. 
Hence the diboson excesses can naturally be explained by the 2 TeV technirho~\footnote{
As noted in Ref.~\cite{Fukano:2015hga}, 
the narrowness reported from the ATLAS group on the 2 TeV resonance (with the width $< 100$ GeV) 
can be eunsured by a suppression factor by $N_F$ characteristic to the one-family model with $N_F=8$, 
compared to the rho meson width in the naive-scale up of QCD with $N_F=2$.  
}. 
 The vector boson model~\cite{Becciolini:2014eba}, on which the diboson analysis in Ref.~\cite{Franzosi:2015zra} has been based,  
can be transformed into the HVT model in Eq.(\ref{LV}). 
 If the (approximate) conformal/scale invariance is present in the underlying theory such as the walking technicolor, 
leading to the effective model in Ref.~\cite{Becciolini:2014eba}, 
 then the matrix of the $V$-$W$-$\phi$ vertices are diagonalized simultaneously with the vector boson mass matrix. 
 Consequently, the same argument as the above becomes applicable to the model in~\cite{Becciolini:2014eba}.

 One way out to avoid the conformal barrier may be to introduce multi Higgs fields which 
give the masses to new vector bosons as well as the weak bosons. 
The mixing among the Higgs bosons would make the mixing structures different 
for the $V$-$W$ and $V$-$W$-$\phi$. 
Models having such a vector boson - Higgs boson sector correspond to those studied in Refs.~\cite{Abe:2015jra,Abe:2015uaa}. 
However, some of those Higgs bosons would phenomenologically be heavy to be integrated out, 
such that, except the lightest 125 GeV Higgs,  
all the Higgs fields in the linear realization can be cast into the nonlinear forms 
keeping only the Nambu-Goldstone boson fields (nonlinear realization). 
The aforementioned models will then be effectively described as a model having the lightest Higgs 
and multi Nambu-Goldstone bosons eaten by weak and new vector bosons (or some of them would be real electroweak pions 
such as technipions).  
Then, the conformal barrier would be operative even for such those multi Higgs models.

In conclusion, 
we have proposed novel handcuffs for new vector bosons in consequence of  
the presence of the (approximate) conformal/scale invariance: 
conformal/scale symmetry prohibits new vector bosons decay to the Higgs.   
The LHC Run-II may probe the presence of the conformal/scale invariance 
hidden in the underlying theory responsible for the existence of new vector bosons: 
conversely, if the decay of new vector bosons into the Higgs 
is not observed in the ongoing  LHC Run-II, then the Higgs can be a dilaton.

\acknowledgments 

We would like to thank Masafumi Kurachi and Koji Terashi for discussions.  
Masaharu Tanabashi is also acknowledged for useful information. 
This work was supported in part by 
the JSPS Grant-in-Aid for Young Scientists (B) \#15K17645 (S.M.).

\end{document}